# Polarization rotation in a ferroelectric BaTiO$_3$ film through low-energy He-implantation

Andreas Herklotz[1,*], Robert Roth[1], Zhi Xiang Chong[2], Liang Luo[2,3], Joong Mok Park[2,3], Matthew Brahlek[4], Jigang Wang[2,3], Kathrin Dörr[1], Thomas Z. Ward[5]

[1] Institute for Physics, Martin-Luther-University Halle-Wittenberg, Halle, Germany
[2] Department of Physics and Astronomy, Iowa State University, Ames, Iowa, USA
[3] Ames National Laboratory, Ames, Iowa, USA
[4] Materials Science and Technology Division, Oak Ridge National Laboratory, Oak Ridge, TN, USA
[5] Center for Nanophase Materials Sciences, Oak Ridge National Laboratory, Oak Ridge, TN, USA

[*] herklotza@gmail.com

**Domain engineering in ferroelectric thin films is crucial for next-generation microelectronic and photonic technologies. Here, a method is demonstrated to precisely control domain configurations in BaTiO$_3$ thin films through low-energy He ion implantation. The approach transforms a mixed ferroelectric domain state with significant in-plane polarization into a uniform out-of-plane tetragonal phase by selectively modifying the strain state in the film's top region. This structural transition significantly improves domain homogeneity and reduces polarization imprint, leading to symmetric ferroelectric switching characteristics. The demonstrated ability to manipulate ferroelectric domains post-growth enables tailored functional properties without compromising the coherently strained bottom interface. The method's compatibility with semiconductor processing and ability to selectively modify specific regions make it particularly promising for practical implementation in integrated devices. This work establishes a versatile approach for strain-mediated domain engineering that could be extended to a wide range of ferroelectric systems, providing new opportunities for memory, sensing, and photonic applications where precise control of polarization states is essential.**

## Introduction

Controlling domain configurations in ferroelectric thin films is one of the most crucial avenues to unlock functionalities in ferroic heterostructures. For example, domain engineering has been shown to be of vital importance in ferroelectric memristors, domain-wall memories and nanocircuitry [1]. New ferroelectric nanoelectronics often rely on the fine tuning and manipulation of domain walls [2] or the creation of complex polar topologies [3], both requiring the control of depolarization fields and lattice anisotropy.

Highly efficient domain engineering thus requires a material system with a high sensitivity towards competing polar and structural phases. A prototype ferroelectric with a multitude of potential domain states is perovskite $BaTiO_3$ (BTO). Bulk BTO is a rhombohedral ferroelectric in its ground state and crosses a series of orthorhombic and tetragonal phases until it reaches cubic symmetry above the Curie temperature of about 110 °C [4]. The extraordinary properties of BTO have made it a cornerstone material for numerous technological applications, from high-density capacitors to piezoelectric actuators. Recently, BTO's strong Pockels effect [5] has drawn significant interest for next-generation electrooptic devices in quantum photonic integrated circuits [6]. While conventional approaches using BTO have demonstrated impressive modulation capabilities for classical optical signals [7], emerging quantum information applications demand precise control over domain configurations to maximize electrooptic coefficients while minimizing optical losses from domain walls and structural inhomogeneities [8]. The ability to deterministically engineer BTO's ferroelectric domain structure could enable high-speed modulators operating at GHz frequencies, essential for quantum information processing, while maintaining compatibility with existing semiconductor fabrication processes. However, achieving such control remains challenging using traditional strain engineering approaches, which are limited by available substrate materials and critical thickness constraints.

An efficient way to engineer domain structures in BTO thin films has been epitaxial strain. Early work has shown that biaxial in-plane strain of only about one percent can shift the ferroelectric transition temperature upwards by several hundred Kelvin [9].

Likewise, domain configurations are effectively altered under biaxial strain. A complex balance of mostly in-plane oriented monoclinic domains in form of a regular zig-zag lamella pattern was observed for a low-strain film grown on $NdScO_3$ substrates [10], while larger compressive strain as grown on $DyScO_3$ or $GdScO_3$ substrates leads to stabilization of the tetragonal phase with predominantly out-of-plane oriented domains [11]. These experimental observations have been in good agreement with phase field simulations based on phenomenological Landau-Ginsburg-Devonshire theory [12,13].

A drawback of epitaxial strain is the need of suitable single crystal substrates to achieve desired strain states. Additionally, strain relaxation occurs above a critical film thickness, which may permit the stabilization of the targeted strain. Alternative approaches to induce lattice strain in a continuous and controllable way may allow access to new domain configurations that are otherwise impossible to achieve by standard strain engineering. In our previous work we have introduced the strain doping approach, where low-energy He ion implantation is used to create a unit cell expansion in thin films [14,15]. In epitaxial thin films this volume expansion is typically equivalent to a uniaxial out-of-plane expansion, since the in-plane lattice remains epitaxially locked to the substrate. The out-of-plane strain can then be continuously controllable via the ion implantation dose and has been used to manipulate magnetic [16,17], dielectric [18,19], and transport properties [20] across a range of oxide films.

In this work we apply strain doping to epitaxial BTO films. We show that the uniaxial strain induced by ion implantation induces a phase transition from a mixed domain to a purely out-of-plane oriented tetragonal domain state. This transition is tantamount to a ferroelectric polarization rotation from in-plane to out-of-plane. We argue that, due to the universal character of strain doping, polarization rotation can also successfully be induced in a wide range of other ferroelectric or ferroelastic systems.

**Results**

A BTO (80nm) / LSMO (10nm) film is deposited on a (110) $DyScO_3$ substrate, capped with 15nm thick Au electrodes and subsequently implanted with 5 keV He ions. A schematic illustration of the heterostructure is presented in **Figure 1a**. The dimensionless He implantation profile, as estimated by a Monte Carlo simulation, is included as a bar graph. The majority of the He ions is expected to stop within the first half of the BTO layer. **Figure 1b** shows a X-ray diffraction (XRD) reciprocal space map around the $(103)_{pc}$ reflections of the as-grown heterostructure. While the LSMO layer is coherently strained to the substrate, the BaTiO$_3$ film shows some degree of strain-relaxation. The pseudocubic lattice parameter of the scandate ($a_{DSO}$ = 3.944 Å) provides a compressive lattice mismatch of -1.2 % to the *a*-parameter of the tetragonal titanate bulk lattice structure ($a_{BTO}$ = 3.992Å, $c_{BTO}$ = 4.036Å). [21–23] Previous work has revealed that epitaxial BaTiO$_3$ films tend to release misfit strain throughout the film growth, leaving a bottom strained part and a strain-relaxed top part [24]. This behavior is reflected in our XRD measurements. The zoomed-in reciprocal space map clearly reveals the presence of a coherently strained film part, represented by a single peak, and a fully relaxed film part, represented by an agglomeration of three sub-peaks. Having a single uniform peak indicates that the bottom film part has tetragonal structure with $a$ = 3.944 Å and $c$ = 4.069 Å and is made up of *c*-oriented ferroelectric domains. This tetragonal strained state will be called the T-phase. The peak splitting of the top part is a consequence of a symmetry change under strain-relaxation. A splitting into three peaks within the H0L plane can be explained by the presence of M$_C$-type monoclinic domains with a polarization vectors within the $(100)_{pc}$ and $(010)_{pc}$ plane. A detailed structural clarification is difficult, since different domain configurations, such as mixed rhombohedral/orthorhombic configurations, can lead to the same XRD peak pattern. Throughout the remainder of the text we will refer to this strain-relaxed domain state as the R-phase, referring to the rhombohedral strain-free bulk-like state. It should be noted that the conclusions throughout the paper are not affected by this uncertainty, since in all cases the crystal symmetry is lower than that of the T-phase and the ferroelectric polarization is rotated away from the film normal.

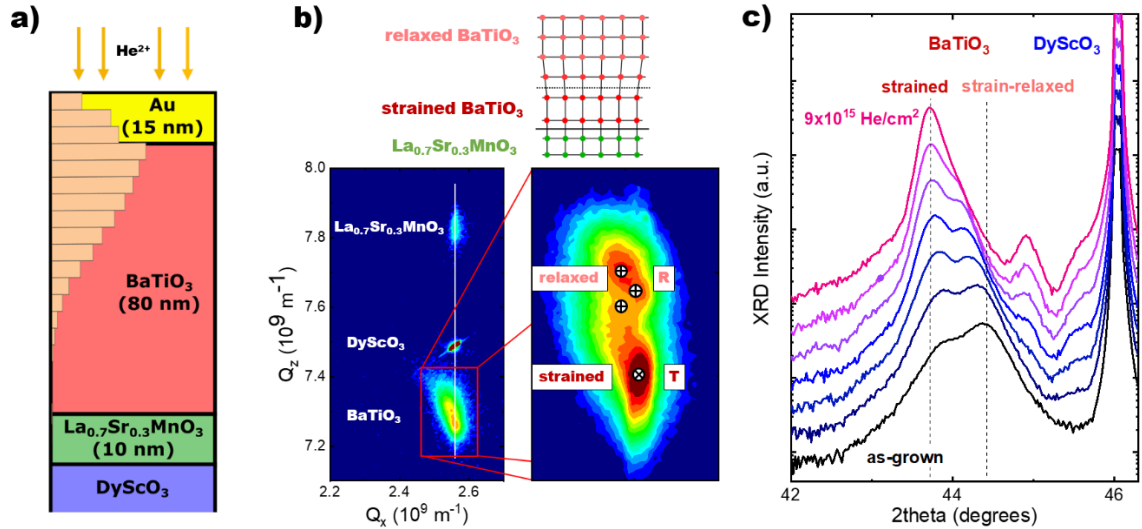

**Figure 1.** a) Schematic representation of the thin film heterostructure studied in this work and the calculated He ion profile. b) A reciprocal space map around the $(103)_{pc}$ reflection highlights partial strain relaxation throughout the BaTiO$_3$ film. The zoomed-in image shows the presence of a strained and a relaxed film part. Crosses indicate peak splitting due to structural domain formation. c) $\theta$-$2\theta$ scans around the $(002)_{pc}$ reflections of the BTO. The scans with varying He implantation levels are shifted vertically for clarity.

**Figure 1c** shows the changes of XRD $\theta$-$2\theta$ scans around the pseudocubic 002 reflections with subsequent He ion implantation. A clear double peak behavior associated to the strained and strain-relaxed film part is visible for the as-grown film. Upon ion implantation the relaxed film peak is shifting to lower $2\theta$ angles to unify with the strained film peak. The natural interpretation of this data is that the top half of the BTO film, which is effectively implanted with He ions, is strained and structurally adapting to the bottom half of the film.

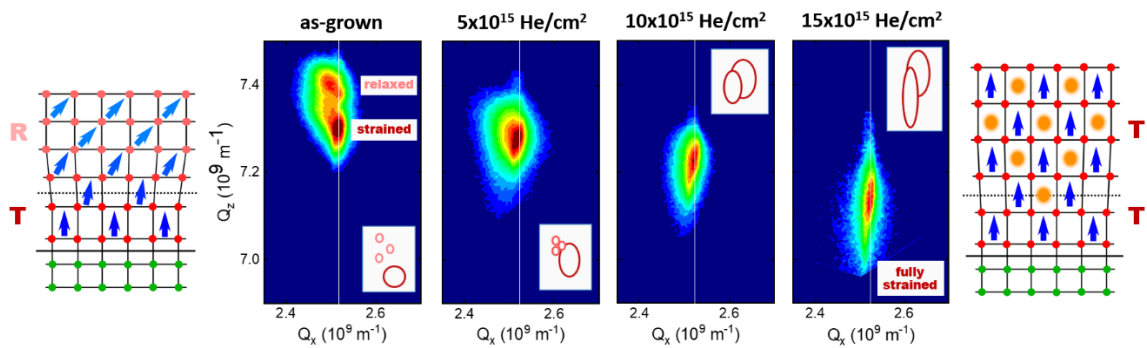

**Figure 2.** Series of reciprocal space maps around the $(103)_{pc}$ reflection of the BaTiO$_3$ film with increasing ion implantation dose. The insets illustrate the positions and shapes of the subpeaks, with the light red circles reflecting the strain-relaxed top part and the dark red circle

reflecting the strained bottom part of the film. The schemes on the left and right illustrate the lattice distortions and polarization direction (blue arrows) of the as-grown and $15\times10^{15}$ He/cm$^2$ film, respectively.

A clearer picture can be derived from reciprocal space maps of the heterostructure taken at various He implantation levels, shown in **Figure 2**. As illustrated schematically on the left, the strained part of the as-grown film has a tetragonal structure with out-of-plane polarization. The strain-relaxed part R-phase part has an average polarization that has a significant in-plane polarization component. Under He implantation the relaxed top part undergoes significant changes. We find that the peak narrows and transforms into a single peak while the film expands along the *c*-axes. We interpret this observation as a strain induced structural transition from the R- to T-state. As illustrated on the right of **Figure 2**, the lattice of the whole film is elongated along out-of-plane and the polarization is fully aligned along the film normal. He implantation effectively transforms the inhomogeneous multi-domain film into a structurally more homogeneous purely tetragonal film.

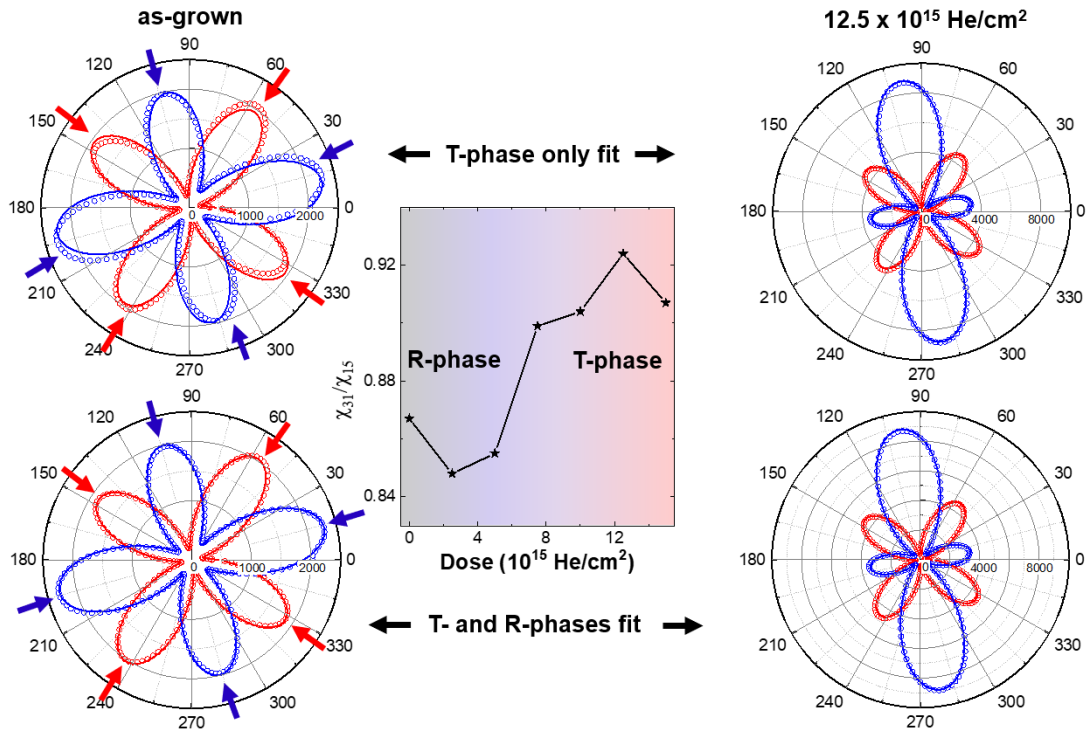

**Figure 3.** SHG measurements: Polar plots of measured s-polarized (red) and p-polarized (blue) SHG intensity as a function of polarization angle φ for the as-grown film and the film implanted with $12.5\times10^{15}$ He/cm$^2$. Measured data and fits are represented by open symbols and solid lines, respectively. Fits are based on the equations for purely tetragonal phase only (top) and a combination of tetragonal and rhombohedral phases (bottom). The red and blue arrows highlight that the as-grown film requires the addition of a rhombohedral phase content to fit the SHG data satisfactorily. The graph in the middle shows $\chi_{31}/\chi_{15}$, as determined from the fitting parameters, as a function of the He dose.

In order to gain additional insight into the symmetry changes induced by He implantation, we have performed second harmonic generation (SHG) measurements on a He implanted BTO heterostructure. **Figure 3** shows polar plots of the measured s-polarized (red) and p-polarized (blue) SHG intensity as a function of the polarization angle φ. The measured data is fit to equations that either reflect a purely tetragonal film or a mix of tetragonal and rhombohedral domains. Detailed information on the experimental setup and derivations of the fitting equations can be found in the supplemental material. When a fully tetragonal crystal structure is assumed for the as-grown film (top left graph), the data can't be fitted satisfactorily. Small but clear deviations can be seen, highlighted by the blue and red arrows. However, a much better fitting result is achieved if the equations for a tetragonal/rhombohedral phase coexistence are used (bottom left). This result is in agreement with our XRD data that indicates that the as-grown film is not fully tetragonal and includes parts with lower symmetry. Note that SHG is sensitive to the whole heterostructure, i.e. to the lower tetragonal coherently strained BTO film part as well as the lower symmetry top part. The film implanted with $12.5 \times 10^{15}$ He/cm$^2$ shows a significantly different SHG response. This change already indicates that He implantation affects the ferroelectric domain structure of the BTO film. The data can be fitted very well by the equations assuming a purely tetragonal film alone (top right graph). Including the terms for a rhombohedral phase does not improve the fit by a notable degree (bottom right graph). This result suggests that He implantation transforms the mixed phase film into a single-phase tetragonal film.

In the center graph of **Figure 3** we plot the second harmonic generation tensor element ratio $\chi_{31}/\chi_{15}$ as a function of He dose. The tensor elements, $\chi_{31}/\chi_{15}$, are determined by extracting SHG intensity at φ =90° in the p-out feature and φ =45° in the s-out feature. An increase of this ratio towards 1 has been shown to indicate an increase of crystal symmetry [25–27]. The overall symmetry of the BTO film is increased during He implantation as the film transforms into a purely tetragonal state. Thus, our SHG measurements serve as a powerful tool for unraveling the spatial symmetry evolution of BTO films. Moreover, as the temporal dynamics of lattice symmetry gain increasing significance [28,29], the SHG findings motivate future time-resolved studies employing ultrafast SHG spectroscopy to investigate the temporal behavior of BTO. In order to corroborate these experimental findings by theory we have performed phase field simulations based on a Landau-Devonshire thermodynamic potential described more in the methods section. At first, the domain configuration as a function of in-plane strain is modelled (first row of **Figure 4**). Under large in-plane compressive strain of -1.2% percent the film consists entirely of dense $c^{+/-}$ domains. This result is in line with previous theoretical and experimental observations [9,12]. It is also in agreement with the absence of XRD peak splitting for the fully strained T-phase film part. When smaller in-plane strain is assumed in our calculations the density of *c* domains decreases. For fully strain free films a large part of the ferroelectric polarization is tilted away from the film normal through the formation of orthorhombic, rhombohedral or monoclinic domains. This predictions has been confirmed experimentally in strain-free films before [10] and is reflected in our data by the observation of the R-phase with splitting into multiple XRD peaks.

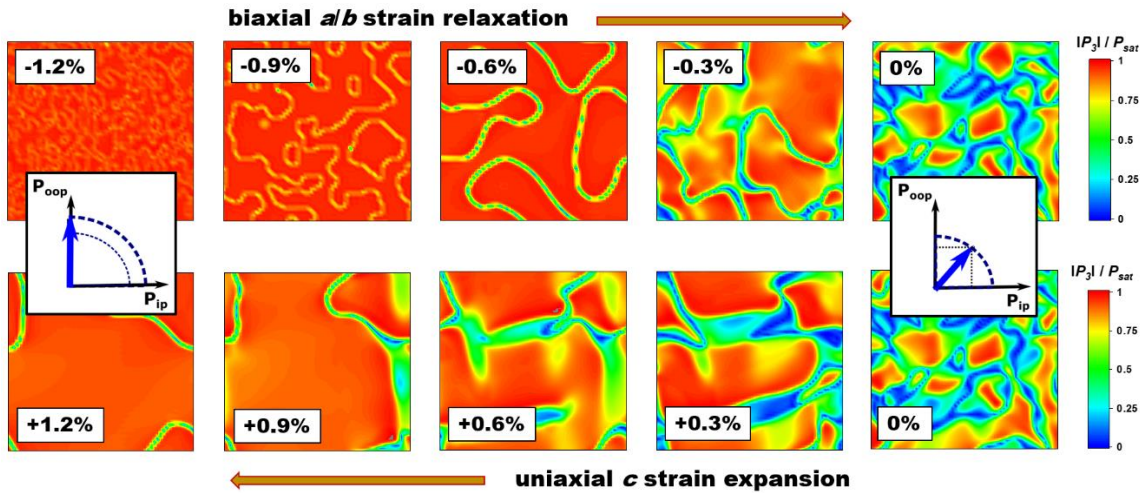

**Figure 4.** The out-of-plane polarization component at the top of a BaTiO$_3$ film as calculated by phase field simulations. The top row highlights the effect of decreasing in-plane strain during strain relaxation, while the bottom row shows the result obtained through continuous expansion of the out-of-plane axis while the in-plane axes is fixed. The insets schematically illustrate the rotation of the average polarization towards the film normal when biaxial compression and out-of-plane expansion is increased.

In a next step, the effect of He implantation on the ferroelectric domain structure was modelled simply by using the strain-free domain configuration as a starting point and sequentially expanding the unit cells along the out-of-plane direction. Second order effects, for example through changing material parameters, were neglected. It can be seen, that the ferroelectric polarization continuously rotates back into the film normal and a $c^{+/-}$ ferroelectric domain pattern is established. The out-of-plane expansion counteracts in-plane strain-relaxation. This result is consistent with our experimental data that suggests a strain-induced structural phase transition of the strain-relaxed part towards a fully tetragonal state.

The structural changes upon He implantation have a profound influence on the BTO thin film properties. During the structural transition towards a fully tetragonal film the uniformity of the film is increased. One might thus expect a reduction of asymmetry in ferroelectric hysteresis loops at lower He doses. In contrast, for higher implantation levels the strain uniformity is decreasing again since ion implantation is only affecting the strain state in the top part of the film, while the bottom part is essentially unchanged. This two-region behavior is confirmed in dielectric hysteresis measurements shown in **Figure 5**. The hysteresis loop for the as-grown film is slightly asymmetric with a shift to negative voltages. This polarization imprint behavior has been observed repeatedly for epitaxial $BaTiO_3$ films and has been attributed to asymmetry within the ferroelectric layer [30], as for example due to accumulation of defects near interfaces, and/or the asymmetry induced by the presence of different top and bottom electrodes [5]. The associated in-built field drives ferroelectric domains to have a favorable orientation, which is upward in our case.

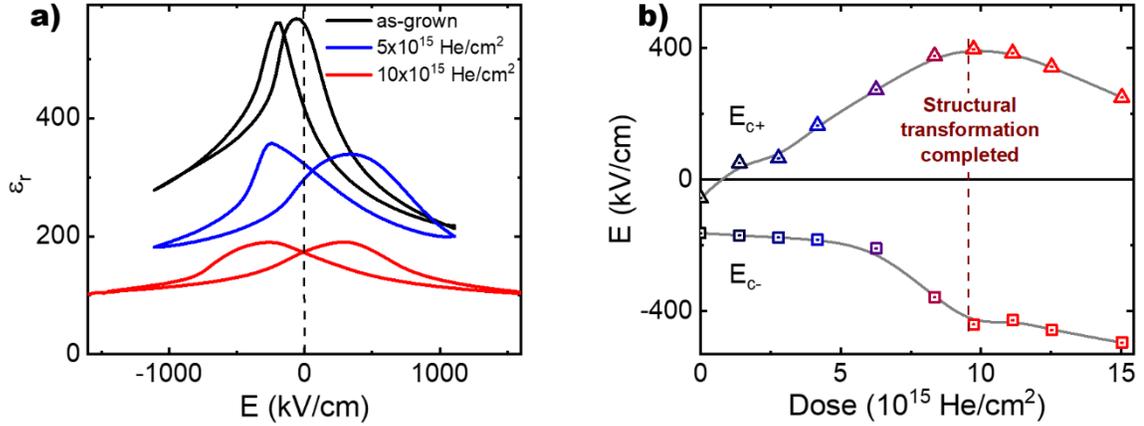

**Figure 5.** a) ε-*E* hysteresis loops for films with different He implantation dose. b) Coercive fields determined from the hysteresis loops as a function of the He dose.

With He implantation the hysteresis loops tend to widen, while the dielectric constants are decreasing overall. Both, enhanced coercive fields and reduced polarizibility, are a consequence of the increased defect density introduced due to ion bombardment and have been reported on other material systems before [32,33]. However, although the coercive fields are increasing, they are doing it at a different rate, as can be seen in **Figure 5b**. At $10 \times 10^{15}$ He/cm² the hysteresis loop is nearly symmetrical. The polarization imprint is removed. This point coincides with the structurally most uniform state where the structural transformation of the top part is fully completed and the out-of-plane strain state is similar to that of the bottom part. The full film is of T-phase. For higher He doses the trend in coercive fields is reversed and an overall negative in-build field is created again. In this region the asymmetry is enhanced again since He ions are only implanted in the top film part and consequently the out-of-plane strain further increases past that of the bottom part. The resultant strain gradient is creating an flexoelectric in-built electric field [30].

**Conclusion**

Helium ion irradiation is demonstrated to offer precise control over ferroelectric domain orientation in BTO films. This approach is found to enable transformation of an as-grown mixed ferroelectric state, characterized by significant in-plane polarization components in the relaxed top portion of the film, into a uniform out-of-plane tetragonal phase. The structural modification significantly improves domain homogeneity while reducing polarization imprint and restoring symmetric ferroelectric switching characteristics. The ability to selectively modify only the top portion of the film through low-energy ion implantation provides a powerful tool for post-growth domain engineering without affecting the coherently strained bottom interface.

Manipulating domain configurations through *ex situ* ion implantation constitutes an appealing strategy to tailor switching characteristics of ferroelectric films. We expect that this technology can be applied to a wide range of material systems with different types of ferroelectric and ferroelastic domain patterns. As an example, the ability to precisely control ferroelectric domains in BTO films through He ion implantation has significant implications for quantum photonic integrated circuits, where efficient electrooptic modulation and low propagation losses are critical. The demonstrated transformation to a uniform out-of-plane polarization state could enhance the Pockels effect needed for high-speed quantum information encoding, while the reduction in domain wall density could minimize optical losses that currently limit device performance. Given that BTO's electrooptic coefficient is approximately 30 times greater than conventional materials like lithium niobate, the ability to optimize its domain configuration could enable modulators operating at GHz frequencies needed for quantum information processing applications. Combined with the technique's CMOS compatibility and room temperature processing, this approach provides a promising pathway toward realizing monolithic quantum photonic devices that integrate efficient modulation with other critical functionalities like photon generation and detection.

**Methods Summary**

*Heterostructure growth*: The BTO/LSMO heterostructure was grown by pulsed laser deposition from stoichiometric targets on a commercial DSO substrate at a deposition temperature of 700°C. The laser fluence for BTO and LSMO was 1.0 and 2.0 kJ/cm$^2$, respectively. The layers were grown in an oxygen pressure of 0.05 and 0.2 mbar, respectively, followed by annealing for 5min and a cool down in 0.2 atm O$_2$.

*He ion implantation*: After film growth, Au films of 15 nm thickness have been deposited on top of the sample to serve as a buffer and neutralization layer for helium ion implantation. The sample was cut into smaller pieces and various Helium doses were implanted using a *SPECS IQE 11/35* ion source at an energy of 5 keV. After implantation, the Au layers were mechanically removed. The ion distribution was simulated with the *SRIM 2013* software package.

*X-ray diffraction*: X-ray diffraction was carried out using a *Panalytical X'Pert* thin film diffractometer with Cu K$_\alpha$ radiation.

*Second harmonic generation*: A laser with 1000 nm wavelength, 150 fs pulse duration and 1 MHz repetition rate was used. The light incident angle is kept constant at 45° with respect to the surface normal. The experimental setup as well as fitting equations are described in more detail in the supplemental material.

*Phase field modelling*: The evolution of the polarization was calculated by solving the time dependent Ginzburg-Landau equations. The framework and all Landau-Devonshire potential parameters were identical to the ones used by Li and Chen. [12] A model size of $64\Delta x \times 64\Delta x \times 48\Delta x$ was employed, with a grid spacing of $\Delta x = 1$ nm. The film and substrate thickness are $30\Delta x$ and $12\Delta x$, respectively.

*Dielectric and ferroelectric properties*: Circular capacitors with a radius of 25 µm were produced by depositing Au electrodes on the heterostructures using magnetron sputtering. Dielectric characterization has been performed with a *HP 4278A* LCR meter.

**Acknowledgements**

This material was based on work supported by the U.S. DOE, Office of Science, Basic Energy Sciences, Materials Science and Engineering Division at Oak Ridge National Laboratory (synthesis and characterization). AH was funded by the German Research Foundation (DFG) - Grant No. HE8737/1-1. The SHG spectroscopy measurement was supported by the US Department of Energy, Office of Basic Energy Science, Division of Materials Sciences and Engineering (Ames National Laboratory is operated for the US Department of Energy by Iowa State University under contract no. DE-AC02-07CH11358). Synthesis and irradiation work was conducted as part of a user project at the Center for Nanophase Materials Sciences (CNMS), which is a US Department of Energy, Office of Science User Facility at Oak Ridge National Laboratory.

## Additional information

## Competing financial interests

The authors declare no competing financial interests.

# Supporting Information for:

# Polarization rotation in a ferroelectric BaTiO$_3$ film through low-energy He-implantation

Andreas Herklotz[1*], Robert Roth[1+], Zhi Xiang Chong[2], Liang Luo[2,3], Joong Mok Park[2,3], Jigang Wang[2,3], Kathrin Dörr[1], Matthew Brahlek[4], Thomas Z. Ward[4,5*]

[1] Institute for Physics, Martin-Luther-University Halle-Wittenberg, Halle, Germany
[2] Department of Physics and Astronomy, Iowa State University, Ames, Iowa, USA
[3] Ames National Laboratory, Ames, Iowa, USA
[4] Materials Science and Technology Division, Oak Ridge National Laboratory, Oak Ridge, TN, USA
[5] Center for Nanophase Materials Sciences, Oak Ridge National Laboratory, Oak Ridge, TN, USA

[*] herklotza@gmail.com
[+] deceased

### Details on the Helium implantation process

Helium implantation was carried out in a vacuum chamber (base pressure 5E-8 Torr) using a *SPECS IQE 11/35* ion source at an energy of 5 keV. An Au films of 15 nm thickness was deposited on top of the BFO thin film before the implantation. This Au layer has several purposes: i) to serve as a buffer layer and prevent sputtering of the BaTiO$_3$ film, ii) reduce the He energy to minimize knock-on damage to the films, iii) neutralize He ions and iv) bring the strain-relaxed part of the BaTiO$_3$ films into the center of the He distribution. The Au layer was removed after the implantation process.

The He ion implantation is carried out continously under a low fluence (~2x10$^{12}$ ions per cm$^2$s). In this low fluence regime sample heating is minimal and doses accumulate linearly, meaning that doubling the dose (i.e. doubling the exposure time to the He ion beam) doubles the He concentration within the lattice and the resulting strain. At the low fluence used, we have not seen any time dependent characteristics. Also, the He doping level remains stable over years if the films are kept near room temperature.

### Second harmonic generation measurements

Second harmonic generation (SHG) experiments we have performed on He implanted BaTiO$_3$/SrRuO$_3$/DyScO$_3$ samples. XRD measurements revealed, that the structural behavior of these films was almost identical to the BaTiO$_3$/LSMO/DyScO$_3$ samples discussed in the main text. We have measured and analyzed the SHG signal of films that are exposed to seven different He doses, an low dose (0 – 3x10$^{15}$ He/cm$^2$), an intermediate dose (3 – 11x10$^{15}$ He/cm$^2$) and a high dose (11 – 15x10$^{15}$ He/cm$^2$).

**Experimental setup.** The geometry of the experiment and the coordinate system is schematically shown below. A light ray of frequency $\omega$ is incident upon the BTO film at an angle $\theta = 45° \pm 5°$ with respect to the surface normal. The in-plane angle $\beta$ with respect to the principal axes of the crystal lattice is unknown in the experiment and used as a fitting parameter below.

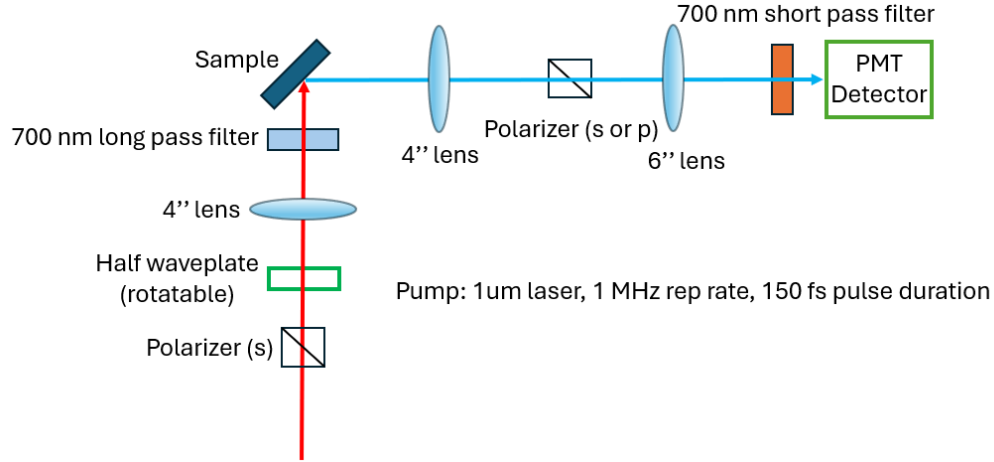

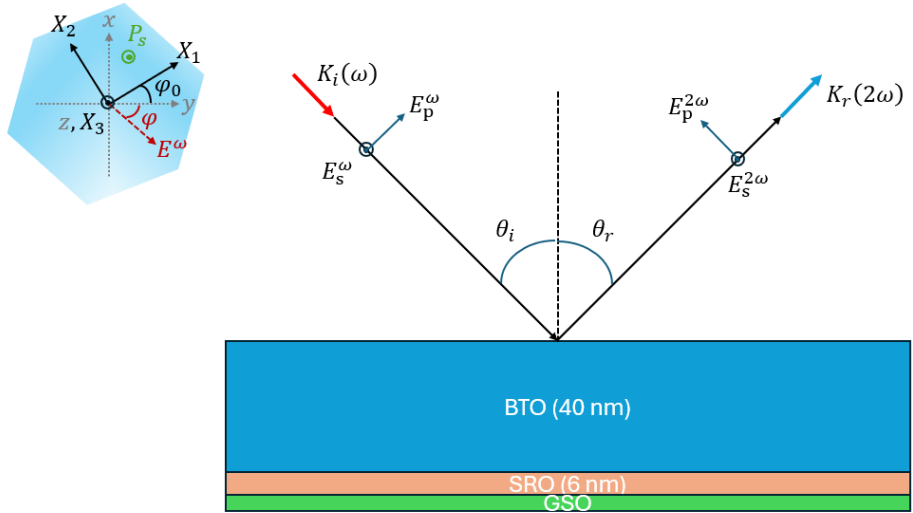

Under these conditions the electric field components of the incoming and outgoing light can be written as follows:

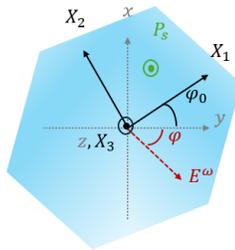

- $E^\omega = (E_1, E_2, E_3)$

- $E_1 = E_0(\cos\phi \cos\theta \cos\phi_0 - \sin\phi \sin\phi_0)$

- $E_2 = E_0(-\cos\phi \cos\theta \sin\phi_0 - \sin\phi \cos\phi_0 \sin\phi_0)$

- $E_3 = E_0(\cos\phi \sin\theta)$

**Note:** $\theta$ stands for the incident angle, set as 45°. $\phi$ stands for the polarization of incident light, which is regulated by a half-wave plate. $\phi_0$ is the angle between lab frame and crystal physics frame.

The field amplitude at the polarizer is given by the second harmonic generation tensor:

$$P_i^{2\omega} = \sum_{j,k} \chi_{ijk}(2\omega;\omega,\omega) E_j^\omega E_k^\omega \qquad E_j^\omega E_k^\omega = E_k^\omega E_j^\omega$$

$$\begin{pmatrix} P_x^{2\omega} \\ P_y^{2\omega} \\ P_z^{2\omega} \end{pmatrix} = \begin{pmatrix} \chi_{xxx} & \chi_{xyy} & \chi_{xzz} & \chi_{xyz} & \chi_{xxz} & \chi_{xxy} \\ \chi_{yxx} & \chi_{yyy} & \chi_{yzz} & \chi_{yyz} & \chi_{yxz} & \chi_{yxy} \\ \chi_{zxx} & \chi_{zyy} & \chi_{zzz} & \chi_{zyz} & \chi_{zxz} & \chi_{zxy} \end{pmatrix} \begin{pmatrix} E_x^{\omega\,2} \\ E_y^{\omega\,2} \\ E_z^{\omega\,2} \\ 2E_y^\omega E_z^\omega \\ 2E_x^\omega E_z^\omega \\ 2E_x^\omega E_y^\omega \end{pmatrix}$$

For the intensity consequently follows:

$$I^{2\omega} = \frac{2\omega^2 |\chi_{eff}|^2 L^2}{n^{2\omega}(n^\omega)^2 c^3 \epsilon_0} \left( \frac{\sin(\Delta k L/2)}{\Delta k L/2} \right)^2 (I^\omega)^2 \qquad \begin{array}{l} I^{2\omega} \propto \left| \chi_{eff}^{(2)} \right|^2 \\ \text{or} \\ I^{2\omega} \propto |P^{2\omega}|^2 \end{array}$$

$$P_i(2\omega) = \epsilon_0 \sum_{j,k} \chi_{ijk}(2\omega;\omega,\omega) E_j^{loc}(\omega) E_k^{loc}(\omega) \qquad E^{loc}(\omega) = \overleftrightarrow{L}(\omega) \cdot E(\omega), \overleftrightarrow{L}(\omega) = L_{xx}^\omega \hat{\imath}\hat{\imath} + L_{yy}^\omega \hat{\jmath}\hat{\jmath} + L_{zz}^\omega \hat{k}\hat{k}$$

$$\chi_{eff} = \langle \overleftrightarrow{L}(2\omega) \cdot \hat{e}(2\omega) \rangle \cdot \frac{P(2\omega)}{2\epsilon_0 (E^\omega)^2}$$

- $L_{xx}^\Omega = \dfrac{2n_1(\Omega) \cos\theta'}{n_2(\Omega)\cos\theta + n_1(\Omega)\cos\theta'}$

- $L_{yy}^\Omega = \dfrac{2n_1(\Omega)\cos\theta}{n_1(\Omega)\cos\theta + n_2(\Omega)\cos\theta'}$

- $L_{zz}^\Omega = \dfrac{2n_2(\Omega)\cos\theta}{n_2(\Omega)\cos\theta + n_1(\Omega)\cos\theta'} \left( \dfrac{n_1(\Omega)}{n_{SHG}(\Omega)} \right)^2$

$n(\lambda)^2 - 1 = \dfrac{4.187\lambda^2}{\lambda^2 - 0.223^2}$

$\lambda$ is in the unit of $\mu m$

For rhombohedral and tetragonal symmetry the SHG tensor reduces:

$$\begin{pmatrix} P_x^{2\omega} \\ P_y^{2\omega} \\ P_z^{2\omega} \end{pmatrix} = 2\epsilon_0 \begin{pmatrix} 0 & 0 & 0 & 0 & \chi_{xxz} & \chi_{xxy} \\ \chi_{yxx} & \chi_{yyy} & 0 & \chi_{yyz} & 0 & 0 \\ \chi_{zxx} & \chi_{zxx} & \chi_{zzz} & 0 & 0 & 0 \end{pmatrix} \begin{pmatrix} (L_{xx}^\omega)^2 E_x(\omega)^2 \\ (L_{yy}^\omega)^2 E_y(\omega)^2 \\ (L_{zz}^\omega)^2 E_z(\omega)^2 \\ 2L_{yy}^\omega L_{zz}^\omega E_y(\omega)E_z(\omega) \\ 2L_{xx}^\omega L_{zz}^\omega E_x(\omega)E_z(\omega) \\ 2L_{xx}^\omega L_{yy}^\omega E_x(\omega)E_y(\omega) \end{pmatrix}$$

$I^{2\omega} \propto \left| \chi_{eff}^{(2)} \right|^2$

$I_p^{2\omega}(\varphi) = \left| \chi_{pp} \cos^2(\varphi + \varphi_0) - \chi_{sp} \sin^2(\varphi + \varphi_0) + \chi_{dp} \sin 2(\varphi + \varphi_0) \right|^2$

$I_s^{2\omega}(\varphi) = \left| \chi_{ps} \cos^2(\varphi + \varphi_0) + \chi_{ss} \sin^2(\varphi + \varphi_0) + \chi_{ds} \sin 2(\varphi + \varphi_0) \right|^2$

$P_x(2\omega) = 2\epsilon_0 \left( 2\chi_{xxz} L_{xx}^\omega L_{zz}^\omega E_x(\omega)E_z(\omega) + 2\chi_{xxy} L_{xx}^\omega L_{yy}^\omega E_x(\omega)E_y(\omega) \right)$

$P_y(2\omega) = 2\epsilon_0 \left( \chi_{yxx}(L_{xx}^\omega)^2 E_x(\omega)^2 + \chi_{yyy}(L_{yy}^\omega)^2 E_y(\omega)^2 + 2\chi_{xxz} L_{yy}^\omega L_{zz}^\omega E_y(\omega)E_z(\omega) \right)$

$P_z(2\omega) = 2\epsilon_0 \left( \chi_{zxx}(L_{xx}^\omega)^2 E_x(\omega)^2 + \chi_{zxx}(L_{yy}^\omega)^2 E_y(\omega)^2 + \chi_{zzz}(L_{zz}^\omega)^2 E_z(\omega)^2 \right)$

We get to:

$$\chi_{eff} = -L_{xx}^{2\omega} \cos\theta \cos\varphi' \begin{bmatrix} 2\chi_{xxz}L_{xx}^{\omega}L_{zz}^{\omega}\sin\theta\cos\theta\cos^2(\varphi+\varphi_0) + \\ 2\chi_{xxy}L_{xx}^{\omega}L_{yy}^{\omega}\cos\theta\sin(\varphi+\varphi_0)\cos(\varphi+\varphi_0) \end{bmatrix} +$$

$$L_{yy}^{2\omega}\sin\varphi' \begin{bmatrix} \chi_{yxx}(L_{xx}^{\omega})^2\cos^2\theta\cos^2(\varphi+\varphi_0) + \chi_{yyy}(L_{yy}^{\omega})^2\sin^2(\varphi+\varphi_0) + \\ 2\chi_{xxz}L_{yy}^{\omega}L_{zz}^{\omega}\sin\theta\sin(\varphi+\varphi_0)\cos(\varphi+\varphi_0) \end{bmatrix} +$$

$$L_{zz}^{2\omega}\sin\theta\cos\varphi' \begin{bmatrix} \chi_{zxx}(L_{xx}^{\omega})^2\cos^2\theta\cos^2(\varphi+\varphi_0) + \chi_{zxx}(L_{yy}^{\omega})^2\sin^2(\varphi+\varphi_0) + \\ \chi_{zzz}(L_{zz}^{\omega})^2\sin^2\theta\cos^2(\varphi+\varphi_0) \end{bmatrix}$$

And with the definition of the following fitting parameters:

$$\chi_{sp} = \chi_{zxx}L_{zz}^{2\omega}(L_{yy}^{\omega})^2 \sin\theta$$

$$\chi_{pp} = \chi_{zzz}L_{zz}^{2\omega}(L_{zz}^{\omega})^2\sin^3\theta - 2\chi_{xxz}L_{xx}^{2\omega}L_{xx}^{\omega}L_{zz}^{\omega}\sin\theta\cos^2\theta + \chi_{zxx}L_{zz}^{2\omega}(L_{xx}^{\omega})^2\sin\theta\cos^2\theta$$

$$\chi_{dp} = \chi_{xxy}L_{xx}^{2\omega}L_{xx}^{\omega}L_{yy}^{\omega}\cos^2\theta$$

$$\chi_{ss} = \chi_{yyy}L_{yy}^{2\omega}(L_{yy}^{\omega})^2$$

$$\chi_{ps} = -\chi_{yxx}L_{yy}^{2\omega}(L_{xx}^{\omega})^2\cos^2\theta$$

$$\chi_{ds} = \chi_{xxz}L_{yy}^{2\omega}L_{yy}^{\omega}L_{zz}^{\omega}\sin\theta$$

We derive at the following fitting equations for tetragonal symmetry:

- $I_s^{2\omega}(\varphi) \propto a + |\chi_{ds}\sin 2(\varphi+\varphi_0)|^2$

- $I_p^{2\omega}(\varphi) \propto b + |\chi_{pp}\cos^2(\varphi+\varphi_0) - \chi_{sp}\sin^2(\varphi+\varphi_0)|^2$

a and b are light polarization independent SHG signal.

And for combined rhombohedral and tetragonal symmetry:

- $I_s^{2\omega}(\varphi) \propto a + |\chi_{ps}\cos^2(\varphi+\varphi_0) + \chi_{ss}\sin^2(\varphi+\varphi_0) + \chi_{ds}\sin 2(\varphi+\varphi_0)|^2$

- $I_p^{2\omega}(\varphi) \propto b + |\chi_{pp}\cos^2(\varphi+\varphi_0) - \chi_{sp}\sin^2(\varphi+\varphi_0) + \chi_{dp}\sin 2(\varphi+\varphi_0)|^2$